\documentclass[aps,pre,twocolumn,showpacs]{revtex4}

\usepackage{graphicx,color,epsfig}

\begin{document}

\title{Homophily, Cultural Drift and the Co-Evolution of Cultural Groups}

\author{Damon Centola$^1$,  Juan Carlos Gonz\'alez-Avella$^2$, V\'{\i}ctor M. Egu\'iluz$^2$ and Maxi San Miguel$^{2}$}

\affiliation{$^1$ The Institute for Quantitative Social Science, Harvard University, Cambridge, MA 02138, USA\\
$^2$IMEDEA (CSIC-UIB), Campus Universitat Illes Balears, E-07122
Palma de Mallorca, Spain}

\date{\today}

\begin{abstract}
In studies of cultural differentiation, the joint mechanisms of
homophily and influence have been able to explain how distinct
cultural groups can form.  While these mechanisms normally lead to
cultural convergence, increased levels of heterogeneity can allow
them to produce global diversity.  However, this emergent cultural
diversity has proven to be unstable in the face of ``cultural
drift"- small errors or innovations that allow cultures to change
from within.  We develop a model of cultural differentiation that
combines the traditional mechanisms of homophily and influence
with a third mechanism of ``network homophily", in which network
structure co-evolves with cultural interaction.  We show that if
social ties are allowed to change with cultural influence, a
complex relationship between heterogeneity and cultural diversity
is revealed, in which increased heterogeneity can reduce cultural
group formation while simultaneously increasing social
connectedness.  Our results show that in certain regions of the
parameter space these co-evolutionary dynamics can lead to
patterns of cultural diversity that are stable in the presence of
cultural drift.
\end{abstract}
%\pacs{89.75.Fb, 87.23.Ge, 05.50.+q}

\maketitle

Homophily - the principle that ``likes attract" - is a prominent
explanation for the persistence of cultural diversity. More
precisely, homophily is the tendency of people with similar traits
(including physical, cultural, and attitudinal characteristics) to
interact with one another more than with people with dissimilar
traits. There are three reasons why homophily is such a powerful
force in cultural dynamics, where "culture" is defined as a set of
individual attributes that are subject to social
influence\cite{1}. Psychologically, we often feel justified in our
opinions when we are surrounded by others who share the same
beliefs - what Lazarsfeld \& Merton\cite{2} call value
homophily\cite{3}\cite{4}; we also feel more comfortable when we
interact with others who share a similar cultural background,
i.e., status homophily\cite{2}\cite{5}\cite{6}\cite{7}\cite{8}.
Both of these reasons are forms of choice
homophily\cite{9}\cite{10}, where patterns of interaction are
driven by preferences for similarity. The third reason, induced
homophily \cite{9}\cite{10}, emerges not from individual choice
but from influence dynamics that makes individuals more similar
over time.

While homophily has been studied empirically as an important
factor in the formation and differentiation of social groups
\cite{7},\cite{8},\cite{11}, there are relatively few formal
models that show how homophily functions to create and preserve
social differentiation \cite{1}\cite{12}\cite{13}. This is because
in addition to the principle of choice homophily, social
scientists also observe the principle of social influence: the
more that people interact with one another, the more similar they
become. This influence process produces induced homophily, in
which the disproportionate interaction of likes with likes may not
be the result of a psychological tendency, but rather the result
of continuous interaction\cite{9}. When choice homophily
(hereafter "homophily") and social influence are taken together,
the explanation of cultural diversity poses something of a
paradox: ``if people who are similar to one another tend to become
more alike in their beliefs, attitudes, and behavior when they
interact, why do not all such differences eventually disappear?"
\cite{1}.

While the processes of homophily and influence can produce global
convergence, Axelrod\cite{1} shows that they can also act as local
convergence mechanisms, which produce emergent social cleavages
that lead to global polarization.  Thus, the answer to the paradox
is that as homophily increases, some groups of people do indeed
converge on their cultural characteristics; yet, if there is
enough heterogeneity in the population, this similarity amongst
group members can also make them even more dissimilar from the
members of other groups\cite{3}\cite{14}.  Ultimately, this can
produce cultural groups that are so dissimilar from one another
that their members cannot interact across group boundaries.  This
shows not only that the simple combination of homophily and social
influence can produce and sustain patterns of global diversity
\cite{1}, but also that the development of cultural barriers
between groups can arise from a process of social
self-organization in which emergent differences becomes
significant enough to prevent inter-group contact, even without
enmity across group lines.

Building on Axelrod\cite{1}, researchers have found that several
factors affect the emergence of cultural diversity, for example
globalization and international communication\cite{15}, cognitive
optimization in social groups\cite{16}, and cultural
drift\cite{17}\cite{18}, i.e., random changes in individual
traits. Cultural drift raises the question of whether the above
explanation of cultural diversity will hold if actors are
permitted to make errors or to develop innovations\cite{1}.
Surprisingly, Klemm et al\cite{17}\cite{18} found that if noise is
introduced at a low rate (allowing cultural traits to change
randomly with a small probability), the basic dynamics of the
homophily and influence model will drive the population away from
cultural diversity and towards cultural
homogeneity\footnote{Kennedy\cite{16} finds a similar result when
homophily is eliminated from the cultural diffusion model.  He
shows that when interactions are not constrained by homophily,
social influence dynamics leads to a homogenous state with a
single global culture.  What is more surprising is that even with
homophily cultural drift will inevitably drive the system to a
global monoculture.}. This happens because the introduction of
random shocks perturb the stability of cultural regions, eroding
the borders between the groups.  This allows the system to find a
dynamical path away from the metastable configuration of
coexisting cultural domains, towards the stable configuration of a
global monoculture\footnote{Klemm et al \cite{17}\cite{18} also
found that if noise acts at high rates, it overwhelms the dynamics
of the model and leads to a state in which distinct cultural
regions never form. In this ``disordered noise regime" cultural
diversity persists, but only as a random pattern of continuously
changing traits.}.  If there is a possibility that small errors or
innovative changes will alter even a few individuals' traits, the
mechanisms of homophily and influence will be unable to sustain
cultural diversity in the long run. Thus, we are faced with the
question of whether global monoculture is an inevitable outcome in
the presence of cultural drift.

The present article takes up this revised form of Axelrod's
question by developing a model that demonstrates conditions under
which local dynamics of homophily and influence can produce and
maintain cultural differentiation even under the noisy conditions
of cultural drift. Other recent attempts to solve the problem of
cultural diversity under drift either fix certain cultural
characteristics\cite{19} or introduce xenophobia into the
dynamics\cite{13}.  We preserve the basic homophily and imitation
dynamics developed by Axelrod\cite{1}.  The key development in our
approach is the specification of homophily.  While both choice and
induced homophily have been primary mechanisms for understanding
how distinct cultural groups can
form\cite{1}\cite{12}\cite{19}\cite{20}, most research in this
tradition places an emphasis upon the changing distribution, or
clustering, of traits over time. However, recent research shows
that network dynamics - the changing patterns of social
interaction over time - may play an equally important role in
understanding the effects of homophily on group
formation\cite{10}. Following this line of research, we introduce
``network homophily" via the co-evolution of individual traits and
network structure\cite{21}\cite{22}\cite{23}. Unlike previous
research on cultural diffusion, the network of social interactions
is not arbitrarily fixed\cite{1}\cite{2}\cite{4}, but rather it
evolves in tandem with the actions of the
individuals\cite{26}\cite{21} as a function of changing cultural
similarities and differences\cite{6}\cite{7}\cite{10}\cite{27}.
Following Klemm et al. \cite{17}\cite{18}\cite{24}\cite{25} we use
the level of heterogeneity in the population as a control
parameter, which allows us to map the space of possible
co-evolutionary outcomes, and thereby to show how network
structure and cultural group formation depend upon one another.
These results allow us to address the question of how stable
cultural groups can be maintained in the presence of cultural
drift.

\section{A Co-evolutionary Model of Cultural Dynamics}

We use an agent-based model\cite{1} in which each actor $i$ has
its individual attributes defined as a vector of $F$ cultural
features; each feature represents a different kind of taste or
behavior (e.g., language, religion, music choice, clothing,...),
and takes its value from a range of $q$ possible traits.  Thus,
the state of an actor i is a vector of $F$ cultural features
$(\sigma_{i1}, \sigma_{i2},\ldots,\sigma_{iF})$, where each
$\sigma_{if}$ corresponds to a cultural trait assigned from the
range of integers between $0$ and $q-1$.  The length of the vector
$F$ represents the social complexity of the population, i.e., the
larger $F$ is, the greater the number of cultural characteristics
that are attributable to each individual\cite{27}.  The number of
traits, $q$, represents the heterogeneity of the
population\cite{27}. The larger $q$ is, the larger the number of
possible traits that a given feature can have, corresponding to a
greater number of cultural options in the society.

The initial state consists of $N$ agents located in a two
dimensional square lattice with von Neumann
neighborhoods\cite{1}\cite{28}. The results of the co-evolutionary
model are not sensitive to the choice of the initial network
configuration, however we begin with the von Neumann lattice
structure for easy comparison with previous
work\cite{1}\cite{17}\cite{18}\cite{29}. Each actor is randomly
assigned $F$ cultural traits. Agents are neighbors if they are
connected by a direct link in the network, where the weight of
this link is determined by their cultural similarity, defined
below.  The dynamics of the model are defined by the following
rules:

\begin{enumerate}
    \item   Select an agent $i$ at random from the population.
    Call $i$ the 'active' agent. From among $i$'s neighbors, select
    a random neighbor $j$ and call this agent the 'partner'.

    \item Calculate the overlap, or cultural similarity, between
    $i$ and $j$ as the number of features on which i and j have the same
    trait: $O(i,j)=\sum_{f=1}^{F}\delta_{\sigma_{if},\sigma_{jf}}$

    \item  If $i$ and $j$ share some features in common, but are not yet identical,
     i.e., $0 < O(i, j) < F$, then $i$ and $j$ interact with probability $O(i, j)/F$.

    \item  Agent $i$ interacts with agent $j$ by choosing a random feature $g$ such
    that $i$ and $j$ do not already overlap, i.e., $\sigma_{ig} \neq \sigma_{jg}$.
    The active agent $i$ then sets its trait at feature $g$ to match its partner's
    trait at feature $g$, i.e.,  $\sigma_{ig}=\sigma_{jg}$.

    \item If $O(i, j) = 0$, $i$ removes $j$ from his network of social ties $T$,
    and randomly selects an agent $k$, where $k \notin T$, $ k \neq j$, $i$  and adds $k$
    to its social network.
\end{enumerate}

Rules 1-4 define the basic homophily and influence model.  Actors
who are similar are more likely to interact.  Further, interaction
makes actors who are similar become even more similar, increasing
the weight of their tie and the likelihood of future interaction.
However, if cultural influence processes create differentiation
between two neighbors such that they no longer have any traits in
common, creating zero overlap in their cultural features and
giving them a tie of zero weight, they still remain fixed as
network neighbors.  Rule 5 incorporates network homophily dynamics
into the model by allowing agents to drop ties to members of their
social networks with whom they have a link of zero weight.
Together, rules 1-5 use the co-evolution of social structure and
individual traits\cite{22} to model the dynamics of cultural
change.

\section{Model Dynamics}

In the absence of co-evolutionary network dynamics (rules 1-4
only), the system can evolve either toward complete homogeneity or
toward cultural diversity, depending on the level of
heterogeneity, $q$, and the number of cultural features,
$F$\cite{1}\cite{29}. In the limit of large $N$, for regular
lattices, random networks and small world networks there is a well
defined transition: for a fixed $F$, there is a critical value of
$q$, $q_c$, that corresponds to the transition from global
monoculture to cultural diversity\cite{24}\cite{25}\cite{29}.
Thus, a single parameter - the heterogeneity in the population -
controls the dynamics of whether the population evolves toward
multiculturalism or toward a global monoculture. This transition
is sharp, going from complete monoculture to widespread diversity
on either side of the critical value\cite{25}. In Figure 1, the
transition is shown by the dramatic change in the average size
(normalized by $N$) of the largest cultural domain, $\langle
S_{max} \rangle/N$, as $q$ increases \footnote{Averages reported
throughout the paper are ensemble averages over 100 realizations
with different random initial conditions.}. When $q < q_c$, the
largest cultural domain approximates the size of the entire
population $\langle S_{max} \rangle \sim N$, indicating little or
no cultural diversity\footnote{$q_c$ is determined as the value of
$q$ for which the fluctuations in the ensemble values of $S_{max}$
is maximum}. However, for $q > q_c$ increased heterogeneity
guarantees that the largest cultural domain is only a small
fraction of the population ($\langle S_{max} \rangle  << N$).
Correspondingly, when $\langle S_{max} \rangle/N$ is small, the
number of distinct cultural groups is large.

%%%%%%%%%%%%%%%%%%%%%%%%%% F1 %%%%%%%%%%%%%%%%%%%%%%%%%%%%%%%%%%%%%%%%%%
\begin{figure}[t]
\includegraphics[width=1.0\linewidth, angle=0]{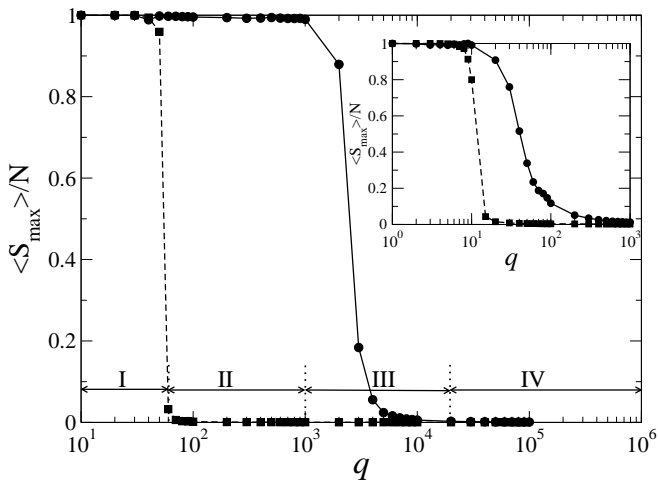}
\caption{Comparison of fixed (squares) and co-evolving (circles)
network models in the transition from global culture to
multiculturalism (no cultural drift included). Results shown are
for $F=10$ and $N=104$ ($q_c ~ 60$). Range of values of q for
regions I-IV are indicated. Inset: Same for $F=3$ and $N = 1024$
($q_c \sim 15$).}
\end{figure}
%%%%%%%%%%%%%%%%%%%%%%%%%%%%%%%%%%%%%%%%%%%%%%%%%%%%%%%%%%%%%%%%%%%%%%%

How does the introduction of co-evolution (rule 5) affect this
transition from global monoculture to multiculturalism? Figure 1
shows that while the qualitative results are the same for fixed
and dynamic networks, introducing network dynamics has the
quantitative effect of increasing the critical value of q. Thus,
there is a large range of values of q for which multiculturalism
is achieved in a fixed network, while co-evolutionary dynamics
lead to a monocultural state\footnote{Figure 1 shows results for
$F=10$ \cite{17}\cite{24} and $F=3$.  Results are qualitatively
similar, but the transition for $F=3$ in a co-evolving network
occurs for lower values of $q$.}.

%%%%%%%%%%%%%%%%%%%%%%%%%% F2 %%%%%%%%%%%%%%%%%%%%%%%%%%%%%%%%%%%%%%%%%%
\begin{figure}[t]
\includegraphics[width=.85\linewidth, angle=270]{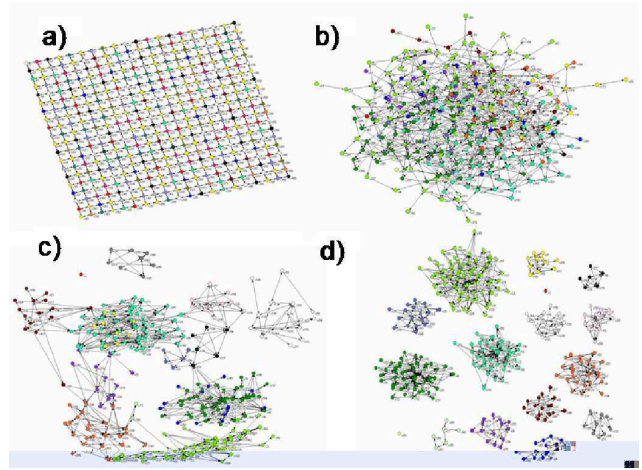}
\caption{Evolution of network, cultural traits and physical groups
in the co-evolving model for $N=400$, $F=3$ and $q=20$. Snapshots
of the network are shown at times a) $0$, b) $2,500$, c) $25,000$,
d) $500,000$.}
\end{figure}
%%%%%%%%%%%%%%%%%%%%%%%%%%%%%%%%%%%%%%%%%%%%%%%%%%%%%%%%%%%%%%%%%%%%%%%

The co-evolutionary dynamics do not only affect the critical value
of $q$, they also dramatically alter the structure of the social
network.  Depending on the value of $q$, the network can evolve
from a regular lattice into a complex connected random network, or
can break apart into multiple components.  This latter point is
quite important, for it means that while cultural diffusion on the
fixed network produces physical boundaries that define the
cultural regions, the dynamic network can self-organize into
culturally distinct physical groups.  This process of
self-organization is illustrated in Figure 2.  Beginning with a
regular lattice (Figure 2a), the system first loses its original
structure (Figure 2b), then forms into culturally homogenous
regions (Figure 2c), which ultimately become culturally homogenous
components (Figure 2d). The colors of the nodes indicate unique
cultural groups, which change over time due to the influence
process.

%%%%%%%%%%%%%%%%%%%%%%%%%% F3 %%%%%%%%%%%%%%%%%%%%%%%%%%%%%%%%%%%%%%%%%%
\begin{figure}[t]
\includegraphics[width=1.0\linewidth, angle=0]{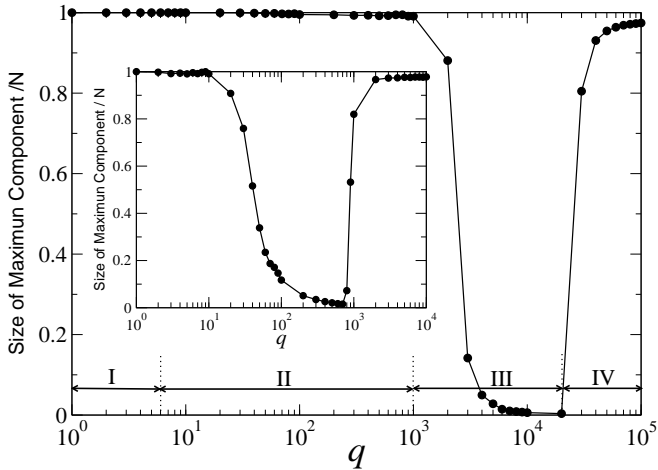}
\caption{Average size of the normalized largest network component
(physical group) in a co-evolving network. Results shown are for
$N = 10^4$ and $F = 10$ $(q^* \sim 2 \times 10^4)$. Inset: Same
for $F=3$ and $N = 1024$ ($q^* \sim 7 \times 10^2)$.}
\end{figure}
%%%%%%%%%%%%%%%%%%%%%%%%%%%%%%%%%%%%%%%%%%%%%%%%%%%%%%%%%%%%%%%%%%%%%%%

These network dynamics are surprisingly dependent upon the level
of heterogeneity in the population, as shown in Figures 3 and 4.
Figure 3 illustrates the effects of increasing $q$ on the size of
the largest network component, or physical group, that is produced
by the co-evolutionary process.  Initially, increasing $q$ causes
component size to decrease; however, for higher levels of $q$
there is a transition above which component size reverses its
trend and increases sharply.  Figure 4 shows the effects of $q$ on
the number of distinct cultural and physical groups.  This figure
also indicates a sharply curvilinear effect of increasing $q$.  In
both figures, these co-evolutionary outcomes are divided into four
distinct regions of the parameter space:

\begin{itemize}
    \item Region I) values of $q < q_c$ ($q_c=60$ for $F=10$ and
      $q_c=15$ for $F=3$) where a global monocultural state is reached in a fixed network.

    \item Region II) values $q_c < q < q_c´$ for which a fixed network attains
     cultural differentiation (shown in Figure 1), while a co-evolving network
     produces a global monocultural state.

    \item Region III) values of $q_c´ < q < q*$ for which multicultural states are produced in
    both a fixed network and in a co-evolving network, and the number of cultural and physical
    groups coincide asymptotically in time in a co-evolving network.

    \item Region IV) values of $q>q^*$, where $q^*$ corresponds to a threshold value
    ($q^* \sim 2 \times 10^4$ for $F=10$, $q* \sim 7 \times 10^2$ for $F=3$) above which the number of cultural
    and physical groups no longer coincide.
\end{itemize}

Region I is the simplest case since there is insufficient
heterogeneity to allow cultural differentiation.  There are too
few cultural options for cultural diversity to be achieved.  In
region II ($60 < q < 10^3$ for $F=10$ and $N=10^4$), there is
sufficient heterogeneity to allow cultural diversity to emerge in
the fixed network.  However, in the co-evolutionary model, the
population remains connected in a single component.  Since actors
in the dynamic network are able to find paths around local borders
by forming new ties, cultural groups break down and the entire
population forms a global monoculture.  As q increases, we
approach region III, in which the dynamic network breaks into
multiple components\footnote{For finite systems (finite $N$), the
transition between regions II and III is not sharply defined and
$q_c´$ is identified with a narrow range of values of $q$ for
which a change in the average co-evolutionary outcome is found.}.
In Figure 3, region III (approximately $10^3 < q < 2 \times 10^4$
for $F=10$ and $N=10^4$) corresponds to values of $q$ for which
there is a gradual decrease in the average size of the largest
physical group.  In Figure 1, this region corresponds to a gradual
decrease in the average size of the largest cultural group.  Thus,
as the size of the largest component decreases, so does the size
of the largest cultural domain.  In Figure 4, this region is also
shown to correspond to the values of $q$ for which there is a
gradual increase of the average number of physical and cultural
groups. So, as the network breaks apart into multiple components,
it also forms into more cultural groups.

%%%%%%%%%%%%%%%%%%%%%%%%%% F4 %%%%%%%%%%%%%%%%%%%%%%%%%%%%%%%%%%%%%%%%%%
\begin{figure}[t]
\includegraphics[width=1.0\linewidth, angle=0]{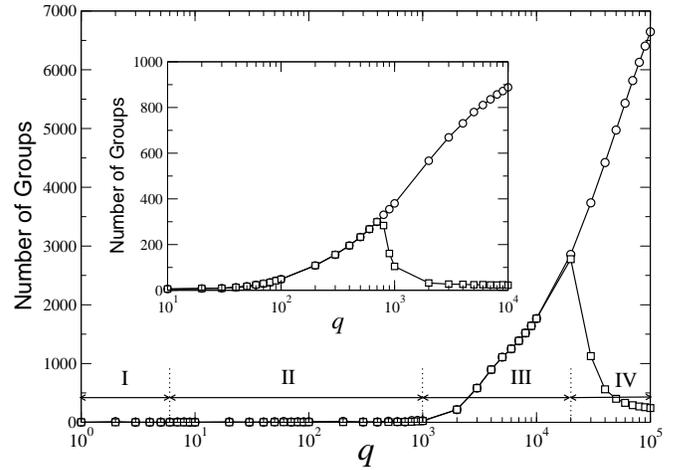}
\caption{Average number of cultural groups (circles) and physical
groups (squares) for $F=10$ and $N=10^4$. Inset: Same for $F=3$
and $N = 1024$.}
\end{figure}
%%%%%%%%%%%%%%%%%%%%%%%%%%%%%%%%%%%%%%%%%%%%%%%%%%%%%%%%%%%%%%%%%%%%%%%

Figure 3 also shows that the trend of decreasing component size,
observed in Region III, is non-monotonic in $q$.  Above $q^*$
(Region IV), the size of the largest component increases sharply.
By contrast, Figure 1 shows no corresponding change in the size of
the largest cultural group, which continues to decrease for $q >
q^*$. Thus, in region IV, the dynamics of cultural group formation
de-couple from the dynamics of network formation.  In regions
I-III the number of cultural groups matches the number of physical
groups, indicating that each component corresponds to a different
cultural domain, however Figure 4 shows that in region IV the
number of cultures continues to increase, while the number of
network components starts to decrease.

Thus, $q^*$ represents a transition in the dynamics of cultural
evolution past which social structure does not determine the
formation of cultural groups.  This is certainly anomalous, since
from the definition of our dynamical model, physical and cultural
groups are expected to coincide asymptotically. Figure 5 sheds
light on this anomalous result by examining the time evolution of
network groups (circles) and cultural groups (squares) for values
of $q$ above (solid) and below (empty) $q^*$.  First, we observe
that both above and below $q^*$, the dynamics of network evolution
(physical group formation) is slower than the dynamics of cultural
group formation. For $q<q^*$, the number of cultural groups (empty
squares) stabilizes at approximately $t=4000$, but the number of
physical groups (empty circles) does not finally converge until
$t=20,000$. For $q>q^*$, the trend is similar, with cultural
groups (solid squares) stabilizing at around $t=20,000$, however
the number of physical groups (solid circles) fails to converge.

This failure of the network to converge highlights the primary
difference between the behavior of the system above and below
$q^*$. As heterogeneity increases, there is an excess of cultural
possibilities, and it becomes less likely that any two actors will
have any traits in common.  Above $q^*$, the large number of
cultural possibilities overwhelms actors in a finite system,
making it difficult for them to find any overlapping traits with
one another. As the size of $q$ becomes of the order $NF$ (system
size times number of features), the number of possible traits is
so much larger than the number of instantiated traits at any given
time that the probability of individuals having any cultural
overlap approaches zero. The consequence is that co-evolutionary
dynamics result in actors continuously breaking links and
searching for new partners in the network, without ever reaching a
stationary configuration.

%%%%%%%%%%%%%%%%%%%%%%%%%% F5 %%%%%%%%%%%%%%%%%%%%%%%%%%%%%%%%%%%%%%%%%%
\begin{figure}[t]
\includegraphics[width=1.0\linewidth, angle=0]{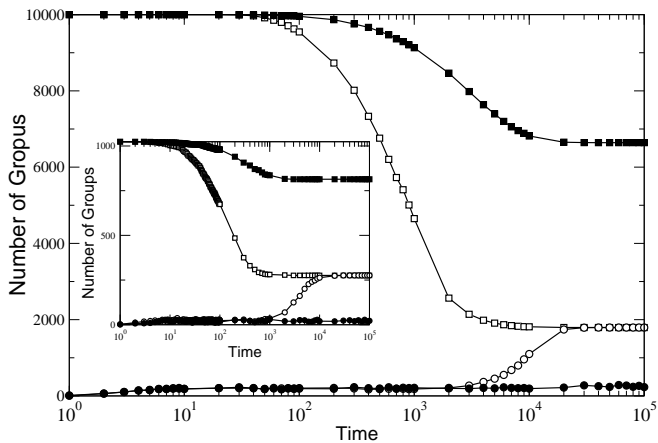}
\caption{Average number of groups as a function of time for $F=10$
and $N=10^4$. Number of cultural groups: empty squares ($q=10^4 <
q^*$), solid squares ($q=10^5 > q^*$). Number of physical groups:
empty circles ($q=10^4$), solid circles ($q=10^5$). Inset: same
for $F=3$ and $N=1024$. Number of cultural groups: empty squares
($q=500 < q^*$), solid squares ($q=6000 >  q^*$). Number of
physical groups: empty circles ($q=500$), solid circles
($q=6000$). }
\end{figure}
%%%%%%%%%%%%%%%%%%%%%%%%%%%%%%%%%%%%%%%%%%%%%%%%%%%%%%%%%%%%%%%%%%%%%%%

Holding the system size and the number of features constant, we
can thus identify how heterogeneity controls the dynamics of
cultural co-evolution. For $q < q_c´$, the social network remains
a single component and the cultural patterns converge on global
monoculture. For  $q_c´ < q < q^*$, the network breaks off into
components that correspond to distinct cultural groups. Finally,
for even greater values of heterogeneity ($q > q^*$), network
evolution and cultural evolution decouple, as the size of the
largest component increases dramatically while cultural groups
fragment into ever smaller - ultimately idiosyncratic - patterns
of traits.

\section{Cultural Drift and Co-evolution}

This analysis of the co-evolutionary dynamics suggests that in the
region of parameter space where nontrivial multicultural states
survive in a co-evolving network, the co-evolutionary cultural
processes of homophily and influence may in fact stabilize the
co-existence of distinct cultural regions even in the presence of
continuous stochasticity.  Following Klemm et al. \cite{17}\cite{
18}, we add cultural drift to the evolutionary dynamics by adding
noise in the form of continuous random shocks, as defined by the
following rule:

\begin{description}
    \item[$6.$] With probability $r$, perform a single feature perturbation.
     A \emph{single feature perturbation} is defined as randomly choosing an agent
     $i$ from the population, $i \in \{ 1,\ldots,N\}$; randomly choosing one of $i$'s features,
     $f \in \{1,\ldots,F\}$; then randomly choosing a trait $s$ from the list of possible traits,
     $s \in \{ 1,\ldots,q\}$, and setting $\sigma_{if}=s$.
\end{description}

Depending on whether the rate of perturbation $r$ is less than or
greater than the time scale on which the homophily and influence
dynamics operate, the system will either be slightly perturbed on
a regular basis (small noise rate), or the system will be
constantly flooded with noise (large noise rate) and unable to
reach any kind of equilibrium. In fixed networks, there is a
critical value of the noise rate $r_c$ above which noise dominates
the behavior of the system\cite{17}. We are here interested in the
small noise rate limit ($r < r_c$), which tests the stability of
cultural diversity in the presence of cultural drift.

As a benchmark for comparison, Figure 6 shows the effects of
cultural drift in region II for a fixed network and for a
co-evolutionary model. For a fixed network (Figure 6a), we observe
that without cultural drift ($r=0$, solid line) the system
stabilizes in a multicultural state   for the whole duration of
the simulation. However, cultural drift ($r=10^{-5}$, dashed line)
drives the system towards a monocultural state,
where\cite{17}\cite{18}. It is worth noting that this monocultural
state is not fixed, as perturbations take the system in random
excursions away from, and then back to, any of the $q^F$
equivalent monocultural states. As a new trait percolates through
the network, the size of the largest cultural group drops as more
people adopt the new trait.  However, as even more people adopt
the trait, the size of the largest group increases again until
cultural uniformity is restored.  For a co-evolving network
(Figure 6b) we observe that after an initial transient the system
orders itself in a monocultural state. This happens in the same
time scale with noise (dashed line) and without noise (solid
line). As in the fixed network, cultural drift causes random
excursions from the final monocultural state, only to return to
another one.

%%%%%%%%%%%%%%%%%%%%%%%%%% F6 %%%%%%%%%%%%%%%%%%%%%%%%%%%%%%%%%%%%%%%%%%
\begin{figure}[t]
\includegraphics[width=1.0\linewidth, angle=0]{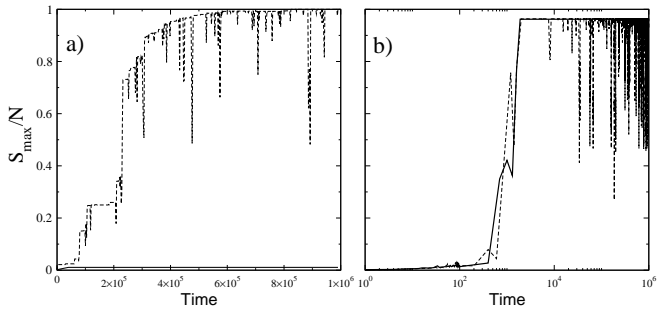}
\caption{: Effect of cultural drift on stability of
multiculturalism in region II. $F=3$, $N=1024$ and $q=20$.
Representative time series for the normalized size of the largest
cultural group ($S_{max}/N$). a) Fixed network. b) Co-evolving
model. Noise parameters: $r=0$ (solid lines), $r=10^{-5}$ (dashed
lines). }
\end{figure}
%%%%%%%%%%%%%%%%%%%%%%%%%%%%%%%%%%%%%%%%%%%%%%%%%%%%%%%%%%%%%%%%%%%%%%%

A more interesting effect is shown in Figures 7a and 7b, which
corresponds to region III. For the fixed network (Figure 7a), the
results are the same as in region II: without noise (solid line)
the system stabilizes with high levels of heterogeneity, but with
noise (dashed line) the system reaches a homogeneous state. As
before, noise-induced excursions away from monoculture give rise
to changes in the cultural make-up of the group, but the system
always returns to a monocultural state.  For the co-evolving
network (Figure 7b), we observe that in the absence of cultural
drift (solid line) the co-evolution model quickly finds a stable
state and then remains in that state for the rest of the
simulation. When cultural drift is added to the co-evolution model
(dashed line), not much happens. The model with noise reaches a
stable state in about the same time, and with $S_{max}/N$ of about
the same size, as it does without noise. Small perturbations
occasionally propagate through the groups, causing shifts in their
cultural identities.  However, the network structure, the number
of physical groups, and the composition of the groups remains
unchanged.  Figure 8 shows the number of cultural groups
corresponding to Fig.7.  As expected, the fixed network without
noise (solid line) stabilizes with a large number of cultural
groups, but when noise is added (dashed line) the number of
cultural groups drops to one.  Conversely, for the co-evolving
network both without noise (solid circles) and with noise (open
circles), diverse cultural groups stabilize in about the same time
and remain in tact throughout the simulation.  While cultural
drift may cause slight changes in the internal culture of the
groups, either through perturbations occurring, then dying out, or
through perturbations successfully propagating through the
cultural groups, the membership of the cultural groups remains
distinct.  Without cross-cutting\cite{27} ties between these
groups, there are no opportunities for new cultural exchanges to
incite cross-border interaction between cultural groups.  Their
isolationism guarantees that they can maintain their cultural
distinctiveness, dynamic though it may be, even in the face of
persistent cultural drift.

%%%%%%%%%%%%%%%%%%%%%%%%%% F7 %%%%%%%%%%%%%%%%%%%%%%%%%%%%%%%%%%%%%%%%%%
\begin{figure}[t]
\includegraphics[width=1.0\linewidth, angle=0]{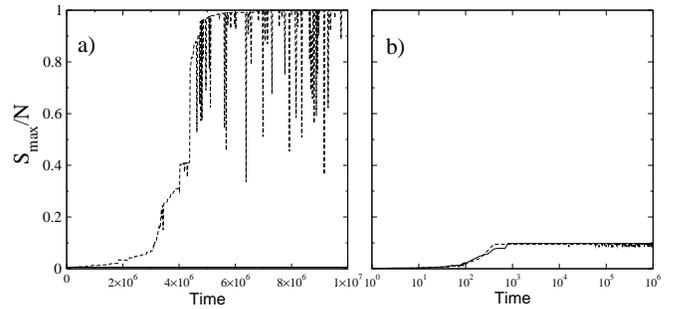}
\caption{Effect of cultural drift on stability of multiculturalism
in region III. $F=3$, $N=1024$ and $q=100$. Representative time
series for the normalized size of the largest cultural group
($S_{max}/N$). a) Fixed network. b) Co-evolving model. Noise
parameters: $r=0$ (solid lines), $r=10^{-5}$ (dashed lines).}
\end{figure}
%%%%%%%%%%%%%%%%%%%%%%%%%%%%%%%%%%%%%%%%%%%%%%%%%%%%%%%%%%%%%%%%%%%%%%%

To understand why cultural drift does not cause cultural groups to
break down, it is necessary to recall that groups will only break
down if they form links to other groups.  However, new links are
only made when existing ties are dropped. Thus, the stability of
groups in the dynamic model hinges on the low likelihood that an
actor will drop a social tie, which is equivalent to the
likelihood of having zero overlap with a fellow group member. Once
groups have formed, the local processes of homophily and influence
create cultural consensus within the group.  Thus, for an actor to
have zero-overlap with one of its neighbors, a sequence of
perturbations must occur such that an actor goes from complete
overlap to zero overlap. A lone perturbation on one feature will
leave the altered actor with a very high level of similarity with
its neighbors. Thus, a single perturbation will result in either
the new cultural feature reverting to its original state (if the
altered actor is influence by its neighbor), or the new cultural
feature being adopted by a neighbor (if the altered actor
influences its neighbor). In both cases, the dynamics of homophily
and influence guarantee that the local group will achieve cultural
consensus on the newly introduced feature, either through its
elimination or its adoption.

In order for similarity between neighbors to decline, an actor
with a new cultural feature must keep the cultural feature without
it either being adopted or eliminated, while a second perturbation
occurs, either to the originally altered actor or to one of its
neighbors. This second perturbation must occur on a separate
cultural feature, and must lessen the overlap between the two
neighbors. Once again, no influence can take place, otherwise
their similarity will increase, leading toward the absorption or
elimination of the new traits. This sequence of perturbations must
occur, without interruption by the processes of local influence, F
times in order for two culturally identical neighbors to develop
zero overlap. The probability of this occurring is roughly
$1/N^F$, or the chance that a single agent will be perturbed $F$
times in a row on a different feature each time. The probability
is even lower if we consider that none of these perturbations can
match any of the neighbors' current traits.  For the systems we
have been studying ($N=10^4$) with $F=10$, the chances of such an
event are less than one in $10^40$.  Furthermore, for the noise
levels used here and elsewhere\cite{17}\cite{18} to represent
cultural drift, the model dynamics operate at a much faster
timescale than do the perturbations (on average, all actors are
activated ten times between each global perturbation), making the
probability that such a sequence of perturbations could occur
before homophily and influence dynamics would recover cultural
consensus infinitesimally small.  Thus, at least during time
scales that are quite large as compared with the timescale of
cultural convergence (approximately $10^3$), multicultural states
in co-evolutionary systems are robust against cultural drift.

%%%%%%%%%%%%%%%%%%%%%%%%%% F8 %%%%%%%%%%%%%%%%%%%%%%%%%%%%%%%%%%%%%%%%%%
\begin{figure}[t]
\includegraphics[width=1.0\linewidth, angle=0]{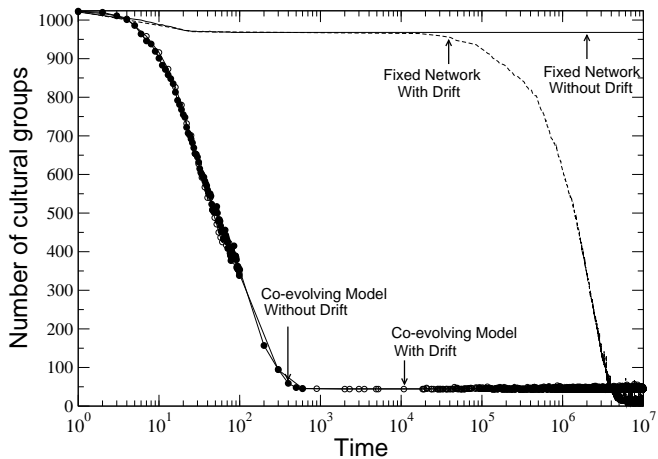}
\caption{Time evolution of number of cultural groups for the time
series of Fig. 7. $F=3$, $N=1024$ and $q=100$. Fixed network
(solid line $r=0$, dashed line $r=10^{-5}$). Co-evolving model
(solid circles $r=0$, open circles $r=10^{-5})$.}
\end{figure}
%%%%%%%%%%%%%%%%%%%%%%%%%%%%%%%%%%%%%%%%%%%%%%%%%%%%%%%%%%%%%%%%%%%%%%%

\section{Conclusion}
Our study of cultural differentiation introduces network homophily
into the dynamics of cultural interaction.  The co-evolution of
network structure and cultural traits reveals a complex
relationship between heterogeneity and the emergence of diverse
cultural groups,  indicating four qualitatively distinct regions
of the parameter space.  In regions I and II, the network remains
connected, and both fixed and co-evolutionary dynamics lead to an
absence of diversity in the presence of noise.  However, in region
III cultural groups can form and stabilize even in the presence of
continuous stochastic shocks.  Physically distinct cultural
components form because increasing diversity creates zero overlap
between neighbors, causing actors to drop these ties and segregate
into discrete cultural groups.  The 'flip side' of this social
detachment process is the formation of social units - homophilous
clusters - that become culturally well-integrated social
structures. "This tendency for network relations to form between
those who have similar social characteristics is known as the
'homophily principle.' Since individuals close to one another on a
dimension of social space are similar, homophily implies that ties
are local in social space." \cite{11}.

In region III, the physical space of the initial lattice network
is rearranged until all ties are "local in social space".  This
process produces an emergent social landscape in which discrete
social clusters correspond to distinct "trait groups". Consistent
with the results of Popalieraz and McPherson\cite{11}, the
interaction of homophily and influence produces a niche structure
whereby peripheral members are either absorbed into the core
beliefs of the social group (by influence), or are forced out of
the social group (by zero overlap).  It is significant, however,
that these social niches are not produced through competition or
selection pressure\cite{11}, but through the mechanisms of
homophily and influence in a co-evolutionary process.  Thus, even
in the absence of selection pressures, a population can
self-organize into stable social niches that define its diverse
cultural possibilities.

As heterogeneity increases, q approaches the threshold at which it
enters region IV.  In region IV, the abundance of cultural options
overwhelms the population, permitting some "anomic"\cite{30}
actors to develop unlikely combinations of cultural features that
prevent them from interacting with anyone.  While some actors are
able to form into homophilous clusters, the anomic actors
perpetually add and drop ties.  When $q>q^*$, the largest
component in the network consists of this disenfranchised group of
actors who are unable to establish memberships in any of the
homophilous social clusters. With increasing heterogeneity, the
number of anomic actors increases, as does the size of this
component, until the entire population forms a single network that
is simply a buzz of adding and dropping ties with no mutual
influence or lasting relationships. These very high levels of
heterogeneity are empirically unrealistic in most cases, however
they warn of a danger that comes with increasing options for
social and cultural differentiation, particularly when the
population is small, or there is modest cultural complexity.
Unlike cultural drift, which causes cultural groups to disappear
through growing cultural consensus, a sudden flood of cultural
options can also cause cultural groups to disappear; but instead
of being due to too few options limiting diversity, it is due to
excessive cultural options creating the emergence of highly
idiosyncratic individuals who cannot form group identifications or
long term social ties. This suggests that, in addition to previous
findings that increased heterogeneity facilitates the maintenance
of cultural diversity, under certain conditions limiting cultural
opportunities may also facilitate the preservation of diverse
cultural groups.

DMC gratefully acknowledges support from the NSF through Cornell's
IGERT program in non-linear dynamics, and through grant
SES-0432917. The rest of us acknowledge financial support from
MEC(Spain) trough project CONOCE2( FIS2004-00953)


\begin{thebibliography}{99}

\bibitem{1}  Axelrod R  (1997)  \emph{J. Conflict Res.}  41:
203-226.

\bibitem{2} Lazarsfeld P and. Merton R K (1954). \emph{Friendship as
a social Process: A Substantive and Methodological Analysis, in
Freedom and Control in Modern Society}, Morroe B, Theodore A, and
Page CH, (Van Nostrand, New York) eds 18-66.

\bibitem{3} Knoke D (1990) \emph{Political Networks: The Structural
Perspective} (Cambridge Univ. Press, N.Y.).

\bibitem{4}  Huston TL, Levinger G  (1978) Ann. \emph{Rev. Psychol.}
29:115-56.

\bibitem{5}  Shrum W, Cheek NH Jr, Hunter SM (1988) \emph{Sociol.
Educ}. 61:227-39.

\bibitem{6} Marsden PV (1987) \emph{Am. Sociol. Rev}. 52:122-313.

\bibitem{7} Marsden PV (1988) \emph{Social Networks} 10:57-76.

\bibitem{8} Fischer CS (1977) \emph{Networks and Places: Social
Relations in the Urban Setting}. (Free Press, N.Y.).

\bibitem{9}  McPherson J M, Smith-Lovin L  (1987)  \emph{Am. Sociol.
Rev.} 52:370-79.

\bibitem{10} McPherson J M, Smith-Lovin L and  Cook J (2001)
\emph{Ann. Rev. Sociol}. 27:415-44.

\bibitem{11} Popielarz P and McPherson JM (1995)  \emph{Am. J. Sociol.
101:698-720}.

\bibitem{12} Mark N. (1998)  \emph{Am. Sociol. Rev. 63:3}.

\bibitem{13} Macy M W, Kitts J Flache A, and Benard S (2003)
\emph{Polarization in Dynamic Networks: A Hopfield Model of
Emergent Structure in Dynamic Social Network Modeling and
Analysis}, (National Academy Press).

\bibitem{14} Duncan OD, Haller A O, Portes A. (1968)  \emph{Am. J. Sociol.
74:119-37}.

\bibitem{15} Greig J M  (2002) \emph{J. Conflict Res}.  46: 225-243.

\bibitem{16} Kennedy J  (1998) \emph{J. Conflict Res}.  42: 56-76.

\bibitem{17} Klemm K, Egu\'{\i}luz V M, Toral R, and San Miguel M (2003).
\emph{Phys. Rev. E} 67, 045101R:1-4.

\bibitem{18} Klemm K,  Egu\'{\i}luz V M, Toral R, and San Miguel M (2005)
\emph{J. Econ. Dyn. Control}  29: 321-334.

\bibitem{19} Durrett R and Levin S A (2005) \emph{J. Econ. Behav. Organ}.
57(3): 267-286.

\bibitem{20} Mark N (2003) \emph{Am. Sociol. Rev}. 68:3-319.

\bibitem{21} Lazer D (2001) \emph{J. Math. Sociol}. 25:69-108.

\bibitem{22} Egu\'{\i}luz V M, Zimmermann M G, Cela-Conde C J, San Miguel M,
(2005) \emph{Am. J. Sociol}. 110:977-1008.

\bibitem{23} McPherson JM, Popielarz P, Drobnic S (1992)  \emph{Am. Sociol.
Rev}. 57:153-70.

\bibitem{24} Klemm K, Eguiluz V  M, Toral R and San Miguel M (2003)
\emph{Phys. Rev. E} 67, 026120:1-6.

\bibitem{25} Klemm K, Eguiluz, V M, Toral R and San Miguel M (2003)
\emph{Physica A} 327:1-5.

\bibitem{26} Macy, M. W. (1991) \emph{Am. J. Sociol}. 97:808-843.

\bibitem{27}. Blau, P. M, and J E Schwartz (1984) \emph{Crossing Social
Circles} (Academic Press, Orlando).

\bibitem{28}. Centola D, Robb W, and Macy M W (2005)  \emph{Am. J.
Sociol.} 110(4):1009-40.

\bibitem{29} Castellano C, Marsili M, and Vespignani A (2000) \emph{Phys. Rev.
Lett.} 85:3536-3539.

\bibitem{30} Durkheim (1897) \emph{Suicide} (The Free Press,N.Y.
reprint , 1997).

\end{thebibliography}
\end{document}